\documentclass[runningheads]{svmult}

\usepackage{makeidx}   
\usepackage{graphicx}  
\usepackage{subeqnar}  
\usepackage{multicol}  
\usepackage{physprbb}  
\makeindex             

\begin{document}
\title*{Cosmological constraint from QSO spatial power spectrum}
%
%
\toctitle{Cosmological constraint from QSO spatial power spectrum}
%
%
\titlerunning{}
%
\author{Kazuhiro Yamamoto
}
\authorrunning{Kazuhiro Yamamoto}
%
%
\institute{Department of Physics, Hiroshima University, 
Higashi-hiroshima, 739-8526, JAPAN
}

\maketitle              

\begin{abstract}
In this paper we consider constraints on the cosmological
density parameters from the spatial power spectrum of QSOs.
We first review an analytic approach to the spatial power 
spectrum of QSOs, then we compare the result of the analytic 
approach with  a preliminary result of the power spectrum 
from the two-degree Field QSO redshift (2QZ) survey. 
From a simple $\chi^2$ test, we show 
that a finite baryon fraction better explains observation 
of the QSO power spectrum, which might suggest a possible 
detection of the baryonic oscillations in the QSO power spectrum.
\end{abstract}

\section{Introduction}

The 2QZ group has recently reported their
preliminary results of the spatial correlation function and the power
spectrum. These analyses are based on an initial sample of 10,000 
QSOs~\cite{yamamotoCroom01,yamamotoHoyle01}. 
Numerical simulation is the most common technique through which
observational results and theoretical predictions can be compared. 
Indeed, the 2QZ group has utilized the Hubble Volume simulation, 
which is a huge N-body simulation of horizon box size, containing 
1 billion mass particles run by the Virgo consortium~\cite{yamamotoFrenk}. 
However, a simple, semi-analytic formula that reproduces numerical 
results would be useful. The purpose of this paper is to report on 
the development of such a semi-analytic formula and to apply it to 
recent observational results. As a demonstration of the usefulness
of our approach, we use the formula to place constraints on the
cosmological density parameters by comparing the theoretical predictions 
to the 2QZ power spectrum reported by Hoyle et al.~\cite{yamamotoHoyle01}.

%
%

\section{Theoretical Formula and a Simple Application}
\newcommand{\cparaky}{c_{\scriptscriptstyle \|}}
\newcommand{\cperpky}{c_{\scriptscriptstyle \bot}}
\newcommand{\bfqparaky}{{ q}_{\scriptscriptstyle \|}}
\newcommand{\bfqperpky}{{\bf q}_{\scriptscriptstyle \bot}}

In the clustering statistics of high-redshift objects in a redshift 
survey, several observational effects must be incorporated for 
careful comparison between theoretical predictions and an observational 
result. 
A useful theoretical formula for the two-point statistics has been 
developed incorporating the redshift distortions due to peculiar 
motion of sources and the light-cone effect simultaneously, as well 
as the geometric distortion~\cite{yamamotoSMY}. 
According to the result, the power spectrum is obtained by averaging 
the local power spectrum $P_0^{\rm a}(k,z)$ over the redshift,
\begin{equation}
  P^{\rm LC}_0(k)={{\int dz W(z) P_0^{\rm a}(k,z)}
             \over {\int dz W(z)}},
\label{yamamotoPcllk}
\end{equation}
with the weight factor 
$W(z)=\bigl({dN/dz}\bigr)^2\bigl({s^2 ds/ dz}\bigr)^{-1}$,
where $dN/dz$ denotes the number count of the objects per unit redshift 
and per unit solid angle, and $s=s(z)$ denotes the distance-redshift 
relation of the radial coordinate that we chose to plot a map of the objects. 
In the expression (\ref{yamamotoPcllk}), $z$-integration arises due to
the light-cone effect within the small-angle approximation, and
the power spectrum $P_0^{\rm a}(k,z)$ is given by 
\begin{eqnarray}
  &&P_0^{\rm a}(k,z)={1\over \cperpky^2\cparaky}\int_0^1 d\mu 
  P_{\rm QSO}\Bigl(\bfqparaky\rightarrow{k\mu\over\cparaky},~
  |\bfqperpky|\rightarrow{k\sqrt{1-\mu^2}\over\cperpky},z\Bigr),
\label{yamamotoscaling}
\end{eqnarray}
where $P_{\rm QSO}(\bfqparaky,|\bfqperpky|,z)$ is the QSO power spectrum,
$\bfqparaky$ ($\bfqperpky$) is the wave number component parallel 
(perpendicular) to the line-of-sight direction in the real space.
In equation (\ref{yamamotoscaling}), with the comoving distance in the 
real space $r(z)$, we defined $\cperpky={r(z)/s(z)}$ and $\cparaky={dr(z)/ds(z)}$.
We model the power spectrum of Q.SO distribution 
by introducing the bias factor $b(z)$,
\begin{eqnarray}
  &&P_{\rm QSO}(\bfqparaky,|\bfqperpky|,z)=b(z)^2
  \Bigl\{1+{\beta(z)}\Bigl({\bfqparaky\over q}\Bigr)^2\Bigr\}^2
  P_{\rm mass}(q,z),
\label{yamamotoPQSO}
\end{eqnarray}
where $q=\sqrt{\bfqparaky^2+|\bfqperpky|^2}$ and we model
the CDM mass power spectrum 
$P_{\rm mass}(q,z)\propto q^nT(q,\Omega_m,\Omega_B,h)^2D_1(z)^2$
with the transfer function $T$ and the linear growth rate $D_1(z)$.
We adopt the fitting formula of the transfer function by 
Eisenstein \& Hu \cite{yamamotoEH}, which is useful when the baryon 
fraction is large.

As a simple application, we consider a cosmological implication comparing with 
the power spectrum from a 
preliminary result of the 2QZ survey. We simply introduce
$\chi^2$ defined by
\begin{eqnarray}
  \chi^2=\sum_{i=1}^{17}{{ \bigl[P^{\rm LC}_0(k_i)-P^{\rm obs}(k_i)\bigr]^2}
                \over{\Delta P(k_i)^2}},
\label{yamamotodefchi}
\end{eqnarray}
where $P^{\rm obs}(k_i)$ is the observational value at $k_i$
and $\Delta P(k_i)^2$ is the variance of observational errors, 
for which we adopt the $17$ data points in the range
$0.012~h{\rm Mpc}^{-1}< k< 0.2~h{\rm Mpc}^{-1}$ 
on Figure 13 in the paper by Hoyle et~al.~\cite{yamamotoHoyle01}.
Figure 1 displays the contours of the $\chi^2$ for various 
cosmological models on the $\Omega_m-\Omega_b/\Omega_m$ plane. 
For the clustering bias, we assumed the form $b(z)={b_0/D_1(z)}$,
where $b_0$ is a constant, which we determined to minimize the value 
of $\chi^2$. Alternation of this assumption does not alter our 
conclusion qualitatively.
From Figure 1 it is clear that the QSO power spectrum favors
the low density universe rather than the standard CDM model 
with $\Omega_m=1$. The minimum of the $\chi^2$ is located at 
$\Omega_m\simeq0.2\sim0.3$ and $\Omega_b/\Omega_m\simeq0.2\sim0.3$. 
An interesting fact is that the QSO power spectrum is better 
explained with the finite baryonic component, though the peak 
of $\chi^2$ is broad and thus the constraint 
is not that tight~\cite{yamamotoMillera,yamamotoPeacock01}.

\begin{figure}[t]
\begin{center}
\includegraphics[width=.6\textwidth]{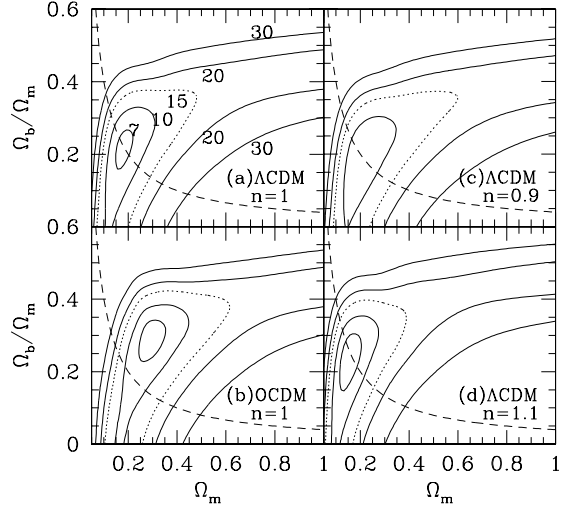}
\end{center}
\caption{Contours of $\chi^2$ on the $\Omega_m-\Omega_b/\Omega_m$
plane for various cosmological models: (a) The $\Lambda$CDM model
with the initial power spectrum with $n=1$;
(b) The open CDM model with $n=1$; (c) The $\Lambda$CDM model 
with $n=0.9$; (d) The same as (c) but with $n=1.1$. 
Here the $17$ data points in Figure 13 in ref.\cite{yamamotoHoyle01} 
are used and $b_0$ is determined to minimize $\chi^2$.
Levels of the contour curves are $\chi^2=7,~10,~15,~20,~30$.
The dotted line is the contour of the level $\chi^2=15$, 
and $\Omega_b=0.04$ on the dashed line.}
\label{yamamotofigcont}
\end{figure}

\section{Conclusion}

By performing a simple $\chi^2$ test we have shown that the QSO power 
spectrum can be consistent with a simply biased mass power spectrum 
based on the familiar CDM cosmology with a cosmological constant. 
We have also shown that the finite baryon fraction better explains observation 
of the 2QZ power spectrum, which might suggest a possible detection of 
the baryonic oscillations in the QSO power spectrum.

\vspace{1mm}
{\it Acknowledgements:}~
The author thanks Fiona Hoyle for providing his results
and also for useful discussions and comments.   
He also thanks Prof. S. D. M. White and 
the people at Max-Planck-Institute for Astrophysics (MPA) 
for their hospitality and useful discussions, where parts 
of this work were done. 
He acknowledges financial supports from the DAAD and the 
JSPS for visiting programs to MPA.

\end{document}